\documentclass[a4paper]{jpconf}
\usepackage{graphicx}
\usepackage[dvips]{hyperref}
\begin{document}
\title{Towards polarizing beam splitters for cold neutrons using superparamagnetic diffraction gratings}

\author{J Klepp$^1$, I Dreven\v{s}ek-Olenik$^{2,3,4}$, S Gyergyek$^4$, C Pruner$^5$, R A Rupp$^1$ and M Fally$^1$}
\address{$^1$University of Vienna, Faculty of Physics, A-1090 Vienna, Austria} 
\address{$^2$University of Ljubljana, Faculty of Mathematics and Physics, SI 1000, Ljubljana, Slovenia}
\address{$^3$Center of Excellence for Polymer Materials and Technologies, SI 1000, Ljubljana, Slovenia}
\address{$^4$Jo\v{z}ef Stefan Institute, Materials Synthesis Department, SI 1001 Ljubljana, Slovenia}
\address{$^5$University of Salzburg, Department of Materials Science and Physics, A-5020 Salzburg, Austria} 

\ead{juergen.klepp@univie.ac.at}

\begin{abstract}
 For holographic gratings recorded in superparamagnetic nanoparticle-polymer composites the diffraction efficiency should -- next to grating spacing, nanoparticle concentration and grating thickness -- depend on the strength of an external magnetic field and the incident neutron spin state. As a consequence, diffraction gratings should be tunable to act as mirrors for one spin state, while being essentially transparent for the other. Thus, polarizing beam splitters for cold neutrons become feasible. 
\end{abstract}

\section{Introduction}
Polarized neutron experiments are a key technique for materials science \cite{WillisBook2009,FurrerBook2009} as well as fundamental physics \cite{Rauch-00,AbelePPNP2008}.  
The most common methods to prepare neutrons in a well-defined spin state parallel or antiparallel to an external magnetic field, i.e. to produce polarized neutron beams, are: \textit{i)} Bragg reflection from magnetic single crystals such as Cu$_2$MnAl Heusler crystals, 
\textit{ii)} total reflection from magnetic mirrors or supermirrors and 
\textit{iii)} spin filters such as polarized $^3$He-gas filters that have high absorption for neutrons with spin antiparallel to the nuclear spins of the gas. The decision for or against one of these three methods depends on the application one has in mind, since all of them have certain advantages and drawbacks. For instance, high degrees of polarization can be obtained with bent polarizing supermirrors, but these devices change the direction of the incident beam and exhibit small-angle scattering. Both properties cause severe problems for some neutron scattering applications. For polarized-SANS, cavity transmission polarizers \cite{KellerNIMA2000,AswalNIMA2008} can be a good option if their length of $\sim\!2$\,m is suitable in a particular setup. In turn, the above problems do not occur when employing $^3$He spin filters. Beam polarizations up to 80\% have been reached \cite{WillisBook2009,FurrerBook2009} with this technique. However, such devices can become a costly alternative due to shortages in $^3$He supplies. 
In the present paper, design of low-cost, business-card size polarizers for cold and very-cold neutrons is proposed.


We employ light-sensitive materials combined with holographic 
techniques to produce diffraction gratings for neutron-optics. The materials exhibit a neutron refractive-index pattern,  
arising from a light-induced redistribution of nanoparticles in a polymer matrix, an effect referred to as the photo (neutron)-
refractive effect \cite{Rupp-prl90,Fally-apb02}:  
Upon illumination with a spatial light pattern, a neutron refractive-index change occurs in the recording material that is proportional to the change of the coherent scattering length density $b_c\rho$. Here, $b_c$ is the coherent scattering length of a specific isotope and $\rho$ is its corresponding number density. 
If superposed light beams generate a periodic modulation of $b_c\rho$, neutron diffraction gratings are recorded. One needs to adjust accessable parameters such as $b_c$, the grating spacing $\Lambda$, the grating thickness $d_0$ and $\rho$, so that peak-values of the diffraction efficiency (or reflectivity) $\eta$ are suitable for a particular application one has in mind.
Depending on $\eta$, diffraction gratings for neutrons can be used as wavelength filters, guides, mirrors or beam splitters for interferometry, for instance. 

Nanoparticle-polymer composites \cite{SuzukiAPL2002} allow efficient tuning of the refractive-index modulation by suitable choice of the 
species of possible nanoparticles, i.e. choice of $b_c$ via the nuclides contained in the nanoparticles. Recently, we have demonstrated that holographic SiO$_2$ nanoparticle-polymer composite gratings can be used as 50:50 beam-splitters \cite{FallyPRL2010} or three-port beam splitters (30:30:30) for cold neutrons \cite{KleppAPL2011}. Exploiting the Pendell\"{o}sung interference effect \cite{Sears-89} by tilting of the gratings \cite{SomenkovSolStComm1978}, diffraction efficiencies of 83\% have been reached for very-cold neutrons \cite{KleppPRA2011}. 

One question appears naturally in view of the above statements: Is it possible to produce holographic gratings whose diffraction properties can be controlled via external means -- apart from tilting -- such as cooling or heating, mechanical strain or electromagnetic fields? The answer is affirmative at least for the latter, as is suggested in the following. 

\section{Magnetic contribution to the neutron refractive index modulation in nanoparticle-polymer gratings} 

A one-dimensional sinusoidal grating
can be described by the periodically modulated refractive index 
\begin{eqnarray}\label{eq:grating}
n(x)=\overline n+\Delta n \cos\left(\frac{2\pi}{\Lambda} x\right),
\end{eqnarray}
where $\overline{n}$ and $\Delta n\simeq\lambda^2\, b_c\Delta\rho/(2\pi)$ are the overall refractive-index change and the refractive-index modulation amplitude, respectively, arising from photon-induced reactions and the resulting redistribution of nanoparticles due to the recording light-pattern. 

It is well known (see, for instance, Refs.\,\cite{Rauch-00} or \cite{Sears-89}) that for a magnetic material the neutron-optical potential $V_n$ -- describing neutron-nucleus interactions -- must be modified by adding a magnetic term $V_m=-\vec\mu\cdot\vec B(\vec r)$ to account for the interaction of the neutron magnetic dipole moment $\vec\mu$ with the magnetic field $\vec B$ inside the material. $V_m$ is related to the magnetic scattering length $b_m(\vec Q)$ that contains the magnetic form factor. 
The neutron refractive index becomes 
\begin{eqnarray}\label{eq:refrInd}
n=\sqrt{1-\frac{V_n+V_m}{E}}\simeq 1-\lambda^2\frac{[b_c\pm b_m(0)]\rho}{2\pi},
\end{eqnarray} 
where $b_m(0)$ refers to 
\begin{eqnarray}\label{eq:bm}
b_m(\vec Q)=
-\frac{m_N}{2\pi\hbar^2\rho}\vec\mu\cdot\vec B(\vec Q)
\end{eqnarray}
at $\vec Q=0$. Here, $E$ and $m_N$ are the energy of the incident neutron and the neutron mass, respectively. 
Since the structure of holographic nanoparticle-polymer gratings is based upon redistribution of nanoparticles, i.e. a modulation of $\rho$, we get 
\begin{eqnarray}\label{eq:DeltaNmagn}
\Delta n\simeq\lambda^2\frac{[b_c\pm b_m(0)]\Delta\rho}{2\pi}
\end{eqnarray} 
for the neutron refractive index modulation amplitude in gratings with \emph{magnetic} nanoparticles. One can see from Eqs.\,(\ref{eq:refrInd}) and (\ref{eq:DeltaNmagn}) that for holographic gratings containing sufficiently high concentrations of magnetic nanoparticles the modulation in Eq.\,(\ref{eq:grating}) can be switched off for one spin component just by adjusting the strength of an external field, i.e. the grating becomes essentially invisible for neutrons in this spin state. Furthermore, by carefully choosing the grating parameters \cite{KleppPRA2011}, the grating can at the same time act as a mirror for the other spin component. That means polarizing beam splitters for cold and very-cold neutrons should be feasible. 

\section{Superparamagnetic nanoparticles}
The superparamagnetic state of noninteracting single-domain ferro- or ferrimagnetic nanoparticles is governed by 
the thermal energy and the potential energy of the particles within an external magnetic field \cite{WagnerBook2005}. Typical relaxation times (the so-called N\'{e}el relaxation time) are very short compared to measurement times and so the net magnetization appears to be zero in the absence of an external magnetic field. On application of a field, the nanoparticles get magnetized as it is the case for paramagnets. However, the magnetization can become very large, depending on the type of nanoparticles and their preparation. 

Next to the micro- or nanostructure of the nanocomposite, mainly the content of nanoparticles in the matrix determines the magnetic properties of the material.
As in most applications, superparamagnetic nanocomposites with high magnetization are desired in our case. It is therefore necessary to prepare a 
nanocomposite containing large amounts of dispersed superparamagnetic nanoparticles. However, to lower their 
surface energy, nanoparticles show a strong 
tendency to agglomerate. 
In order to prevent such agglomeration and to provide for 
compatibility of nanoparticles with the polymer matrix it is essential to coat the nanoparticles with a suitable surfactant (see, for instance, \cite{GyergyekColl&Surf2008,MakovecColl&Surf2009}). 
\begin{figure} \center
\includegraphics[width=100mm]{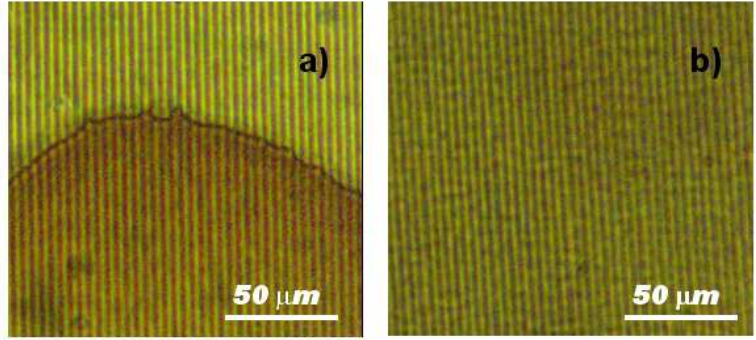}
       \vspace{0.5ex}
       \caption{ \label{fig1}
Optical polarization microscopy image of grating structures with maghemite nanoparticles. Dark regions are rich with maghemite.
a) Structure before secondary polymerization, b) after secondary polymerization. }
      \end{figure}

Development of optical methods to structurize superparamagnetic nanocomposite materials as described in the latter two references are currently under way. Preliminary results for maghemite nanoparticles are shown in Fig.\,\ref{fig1}. 
Many of the superparamagnetic nanoparticle species absorb optical radiation. This brings about some problems to usual methods of holographic structuring. 
Therefore, a two-step structuring method is considered. First, a template grating structure is recorded in transparent materials. Second, one of the contituents of the template is removed by a suitable solvent and the remaining voids are refilled with the solution of magnetic nanoparticles in a photosensitive monomer. After refilling, secondary photopolymerization by homogeneous illumination ensures that the final composite structure remains stable. Note that estimations of nuclear and magnetic (saturation) scattering length densities of maghemite nanoparticles are $b_c\rho\simeq 6.7\times 10^{10}$cm$^{-2}$ and $b_m\rho\simeq 1\times 10^{10}$cm$^{-2}$, respectively (see, for instance, \cite{AvdeevJApplCryst2009} and references therein). 
Materials with larger $b_m/b_c$ will be investigated in the near future.

\section{Conclusion} 

In this paper, small sized and low-cost polarizing beam splitters for cold and very cold neutrons -- based upon diffraction gratings recorded in superparamagnetic nanoparticle-polymer composites by holographic means -- have been proposed. Because of the spin-dependent magnetic contribution to the coherent scattering length, the properties of such diffraction gratings can be tuned such that for one spin component the grating acts as a mirror, while for the other it is transparent. Due to high absorption of maghemite nanoparticles in the visible wavelength range, a new method to structurize such materials using optical holography is currently developed.

\ack
Financial support by the Austrian Science Fund (FWF): P-20265 and the \"{O}AD in the frame of the bilateral project SI 07-2011 is greatly acknowledged. 

\section*{References}
\bibliography{C:/Users/juergen/data/LaTex/localtexmf/bibtex/bib/juergen}

\end{document}